\documentclass[a4paper,twocolumn,rsi]{revtex4}
\usepackage{graphicx}
\begin{document}

\title{ Plasma diagnostics using digital holographic interferometry}
\author{Joseph Thomas Andrews}
\email[Corresponding author: ]{jtandrews@sgsits.ac.in, www.sgsits.ac.in}
\affiliation{Department of Applied Physics, Shri G S Institute of Technology
\& Science,  Indore - 452 003, India. }
\author{Kingshuk Bose}
\affiliation{Department of Applied Physics, Birla Institute of Technology, Mesra, Ranchi - 835 215, India. }
\begin{abstract}
The advances in Charge Coupled Devices in one hand and the high resolution measurements of holographic technique on the other hand, we have adopted the method of digital real-time holographic interferometry  for the diagnostics of  high density plasma. The measured values of plasma electron density agree
        with the measurements from other techniques. 
\end{abstract}
%\ocis{280.5395, 280.3420,  090.0090}
\maketitle

%\section{Introduction}

 In recent years, digital holographic interferometry and shadowgraphy are
 emerging as a versatile tool for non destructive testing and diagnostics purposes
 in materials engineering \cite{recent}. Similar techniques are also adopted
for diagnosing plasma and to measure various plasma parameters using pulsed lasers. However a plasma diagnostics method that can reliably and rapidly measure fluctuations with both spatial and time resolution can lead to increased understanding and eventual control of turbulent transport. Additionally, reliable diagnostics that measure the plasma profiles are critical to moving towards an actual prototype magnetic fusion reactor. Conventional methods for
plasma diagnostics include \cite{paul} Langmuir probe \cite{Dude}, microwave, laser interferometry \cite{Weber} and Thompson scattering \cite{Weber,Game}, Out of these, laser interferometry has the advantage as it is versatile and gives  accurate results. Since the advent of optical lasers, interferometry has been widely used to study different types of plasmas \cite{LPP}. We have made an attempt to successfully adopt the digital holographic interferometry for plasma diagnostics.  To  authors knowledge, this is the first time digital holographic interferometry is employed for plasma discharge tubes in real-time. This open up the way for the diagnosis of arc plasma, DC discharge plasma, etc. The advantages of this technique over other conventional spectroscopic methods are its high resolution measurements of the order of few tens of nanometer, real-time measurements and cost effectiveness.
Further this methods could give realistic parameters of plasma like electron density and electron temperature \cite{recent}.  

The present technique provides information in the form of two-dimensional maps of the electron density, without the need of extensive modeling.
In a laser interferometry various types of interferometer can be used such as Michelson interferometer, Mach-Zhender interferometer (MZI) etc. We employed a Mach-Zhender interferometer. Chord integrated phase information is utilized to obtain the radial profile of electron density by using Abel inversion technique with appropriate software code written using MathCAD$\texttrademark$  and LabVIEW$\texttrademark$. 

%\section{Theory}

We adopt the fact that the geometrical and optical lengths traveled by light through plasma are different, since the refractive index of plasma is proportional to the density of free electrons. When a laser beam passes through a cylindrical tube containing circular symmetric plasma density $(n_e$), the light suffers a change in optical path length. 
Light passing through different chords of the tube as shown in Fig. 1, are phase shifted by different amount. The change in phase is estimated from
\cite{book,Lisi}
\begin{figure}[ht]
\begin{center}
\includegraphics[width=6cm,]{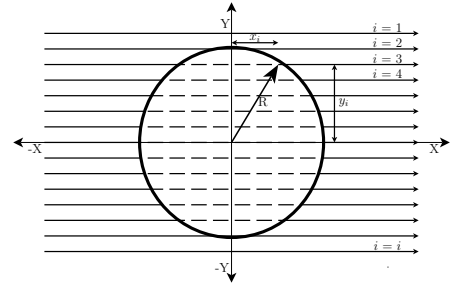}
\end{center}
\caption{Schematic of light propagation inside plasma tube of radius $R$.
The solid horizontal arrows parallel to $x$ axis represents laser beams. The geometrical path length traveled by light inside plasma tube at $i$th chord is $2 x_i = 2 \sqrt{R^2 - y_i^2}$.}
\end{figure}

\begin{eqnarray}
\Delta \phi(y_i) &=&\Delta k \cdot x_{i} \nonumber \\
&=&2\left({ 1- \frac{\lambda_{0}e^{2}\Delta n_{e}(y_{i})}{4\pi c^{2}m_{e}\epsilon_{0}}}\right) \sqrt{R^{2}-y_{i}^{2}},
\end{eqnarray}
where $y_i$ is the horizontal distance of $i$th chord from origin, $\lambda_{0}$, $e$, $\Delta n_{e}(y_i)$, $c$, $m_e$ and $\epsilon_0$ are the wavelength of laser light, charge of an electron, change in electron plasma density at $i$th chord, speed of light, mass of electron and absolute permittivity, respectively. Free electrons produce  negative phase shift because the effective index of refraction  is less than 1. Hence, the change in refractive index will be less than 1, but positive. 

%\section{Experiment}
The experimental setup is shown in Fig. 1. A 30 mW polarized He-Ne laser (Melles Griot) having central wavelength at 632.8 nm passes through a spatial filter assembly and a beam expander (BE). A cylindrical lens is used to convert
the circular symmetric beam into a line beam. The line helps us to measure
the optical properties along a single cross section of the plasma tube only.
Also, the plasma tube, if mounted on a linear stage may be useful to scan
the plasma tube completely.  The expanded beam is divided using a 50:50 non-polarizing cubic beam splitter (BS1). A part of the beam is allowed to incident on a plane mirror (hereafter, we call it reference beam) while the other part is reflected by another mirror (object beam). In this experimental setup, \begin{figure}[ht]
\begin{center}
\includegraphics[width=8cm,]{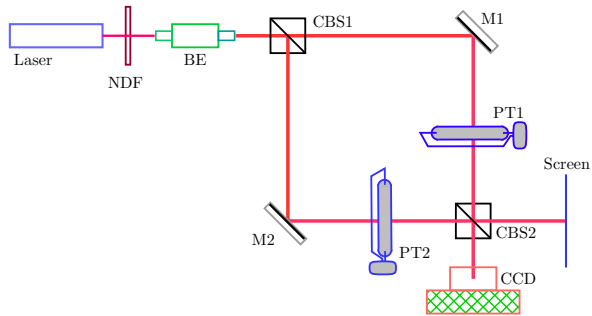}
\end{center}
\caption{Schematic of experimental setup used for digital holographic interferometry. NDF - Neutral density filter, BE -  beam expander, CBS1,2 - cubic beam splitters, M1,2 - mirrors, PT1,2 - high pressure Hg-plasma tubes, CCD - charge coupled device.}
\end{figure}
we have used high pressure mercury lamps (Phillips) as a plasma source. Two identical
lamps are used in both the arms of the Mach-Zhender setup. One lamp acts as a compensator while the other lamp is connected to   power source.  Both the beams interfere inside a non-polarizing cubic beam splitter (BS2). 
Interference fringes are monitored using  a charge coupled device (CCD) (Apogee, Model- LISAA-M). A large area convex lens assembly (Pentax) is used to collect the beam into the CCD. The whole system is mounted on a vibration isolation table (Melles Griot). 
Interference fringes are viewed from one arm of BS2 using a screen. In the absence of any plasma, the interference fringes (or chord integrated intensity) recorded in the CCD can be expressed as~\cite{Born} 
\begin{equation}
I(y_i)= I_1+I_2+ 2 \sqrt{I_{1}I_{2}}\cos( \phi ). \label{I0}
\end{equation}
Here, $I_1$ and $I_2$ are the intensity of the laser in the two arms of the interferometer while $\phi(t)$ is the phase difference between the two beams. Since, the two beams are passing through a path which is identical but intentionally at a small angle so as to obtain straight line fringes. In the presence of plasma eq. (\ref{I0}), may be rewritten as
\begin{equation}
I(y_i, t)= I_1+I_2+ 2 \sqrt{I_{1}I_{2}}\cos[ \phi(t)+\Delta\phi(y_i,t)]. \label{IP}
\end{equation}

%\section{Results and Discussions}

Simulated results obtained using eq. (\ref{IP}) and experimental observation of chord
integrated intensity  are shown as contour plots  in Figs. 3a and 3b, respectively.
Due to the limitation of the experimental setup we could  obtain
only a partial image from the CCD. A typical interference signal obtained using CCD is shown as Fig. 3c. The fringes continuously obtained with time are displayed
in Fig. 3b.  While obtaining the simulated results, the values  of constants
are obtained from the present experimental conditions. The
experimental data is obtained from CCD, while the simulated results are generated
from a software code written using LabVIEW.  For times $t = 0$ to 50 sec no current is applied to PT1. Since no temporal change in phase is occurring, the fringe pattern remains almost same as evident from Figs. 3a and 3b. At $t = 50$sec, constant current is applied to the plasma
source. In the presence of plasma the interference pattern are modulated \begin{figure}[htb]
\begin{center}
\includegraphics[width=8cm,]{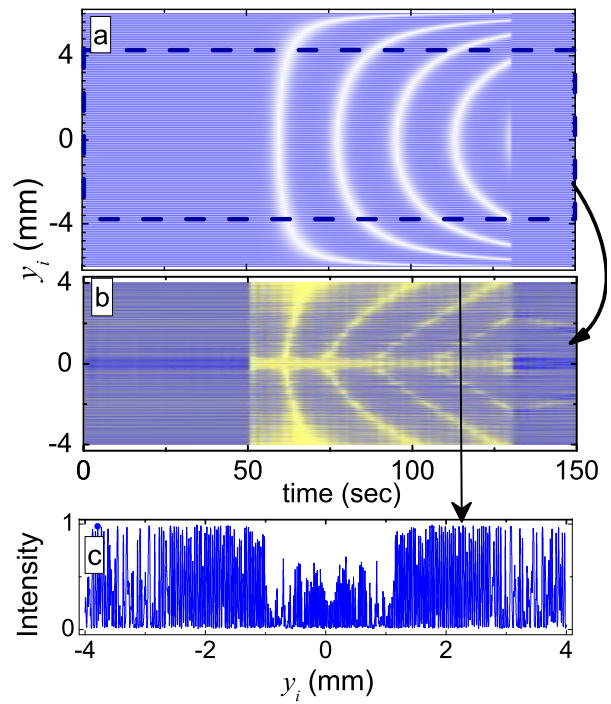}
\end{center}
\vspace{-5mm}
\caption{(a) Contour plot of  simulated results chord integrated intensity
with time and spatial positions. (b)Contour plot of experimental results of chord integrated intensity with time. (c) Chord integrated intensity at $t$ = 118sec is shown with pixel positions. In all the figures, the position of chord is represented as $y_i$.}
\end{figure}
with time. The number of fringes increases due to large change in plasma density, but saturates after the build-up (rising) time of plasma. The unknown parameter  $\Delta n_e(y_i)$ for different chords is obtained after performing many iterations of the simulated results for minimum standard deviation. Best profile of $\Delta n_e(y_i)$ which matches gives minimum standard deviation with the experimental profile is used for calculation. 

The measurement of phase change indirectly gives the magnitude of electron
density. Figure 4, shows the temporal change in plasma density. When the
plasma tube is switched ON, the electron density raises rapidly and saturates.
However, it cools slowly after switching it OFF.

\begin{figure}[htb]
\begin{center}
\includegraphics[width=8cm]{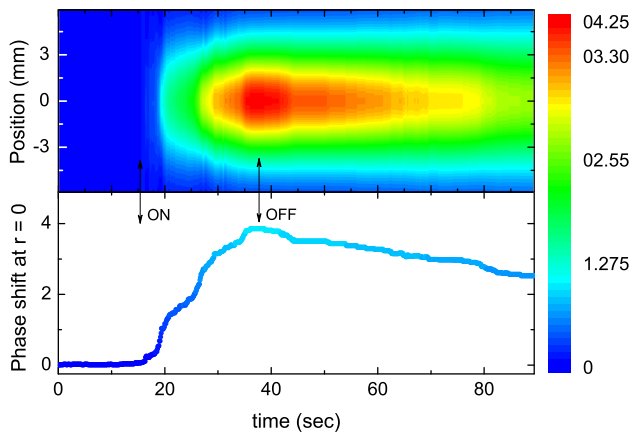}
\end{center}
\vspace{-0.5cm}
\caption{Contour plot of change in phase with axial position and time. The
plasma tube is switched ON at 18th second and switched off at 35th second
and allowed cool till 90sec. Maximum phase change of 4$\pi$ is recorded.
The bottom curve is data at the center ($r=0$mm).} 
\end{figure}

\begin{figure}[htb]
\centerline{\includegraphics[width=8cm]{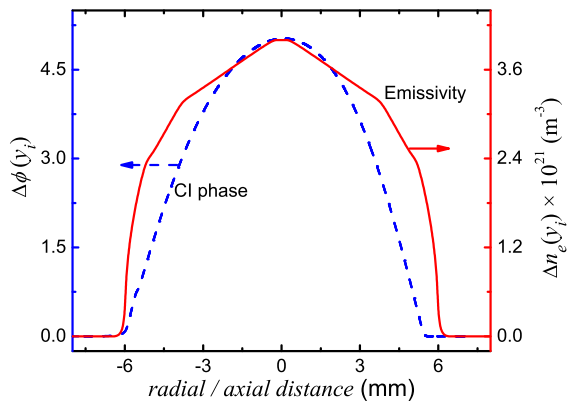}}
\vspace{-0.5cm}
\caption{Experimental measurement of phase change (obtained from chord integrated
intensity) with
axial distance and
electron density (estimated from emissivity profile obtained after Abel inversion process) with radial distance.}
\end{figure}
The chord integrated phase changes with respect to axial distance ($y_i$) are converted radial distance by using Abel inversion method and are shown
in Figure 5. The axis on the left shows the numerical values of phase change
while the axis on the right side is the estimated values of change in electron
density after Abel inversion. These values agree with  the recent observations using spectroscopic method \cite{Nima}. 

%\section{Conclusion}
To conclude, we used the method of holographic interferometry to estimate the plasma
electron density. The measured value agrees with the standard values of high pressure mercury plasma density. The present work has the potential of measuring, imaging the plasma density variations in real-time. This method can also be used for obtaining the 3D profile of electron density in a DC plasma.

\acknowledgements{The authors thankfully acknowledge the financial support
received from } DST\ \& AICTE, New Delhi. The authors also  thank Professors P. K. Sen and   P. K. Barhai for Discussions.

\end{document}